\documentclass[review]{elsarticle}

\usepackage{lineno,hyperref}
\modulolinenumbers[5]

\journal{Journal of \LaTeX\ Templates}

\usepackage{graphicx}
\graphicspath{{Figures/}}
\usepackage{verbatim}
\usepackage{gensymb}
\usepackage{bm}
\usepackage{color}

\begin{document}

\title{Global transition path search for dislocation formation in Ge on Si(001)}

\author[COMP,Aalto]{E. Maras}
\author[Oleg]{O. Trushin}
\author[Alex]{A. Stukowski}
\author[COMP,Aalto,Brown]{T. Ala-Nissila}
\author[Aalto,Iceland]{H. J\'onsson}

\address[COMP]{COMP Center of Excellence, Department of Applied Physics, FI-00076 Aalto, Finland}
\address[Aalto]{Aalto University School of Science, FI-00076 Aalto, Espoo, Finland}
\address[Oleg]{Institute of Physics and Technology, Yaroslavl Branch, Academy of Sciences of Russia, Yaroslavl 150007, Russia}
\address[Alex]{Institut f\"ur Materialwissenschaft, Technische Universit\"at Darmstadt, D-64287 Darmstadt, Germany}
\address[Brown]{Department of Physics, Box 1843, Brown University,
Providence, RI 02912-1843, U.S.A.}
\address[Iceland]{Faculty of Physical Sciences, University of Iceland, 107
Reykjav\'{\i}k, Iceland}


\begin{abstract}
Global optimization of transition paths in complex atomic scale systems is addressed in the context
of misfit dislocation formation in a strained Ge film on Si(001).  Such paths contain multiple intermediate minima
connected by minimum energy paths on the energy surface emerging from the atomic interactions in the system.
The challenge is to find which intermediate states to include and to construct a path going through these intermediates
in such a way that the overall activation energy for the transition is minimal.
In the numerical approach presented here, 
intermediate minima are constructed by heredity transformations of known minimum energy structures 
and by identifying local minima in minimum energy paths
calculated using a modified version of the nudged elastic band method.
Several mechanisms for the formation of a 90$\degree$ misfit dislocation at the Ge-Si interface are identified 
when this method is used to construct transition paths
connecting a homogeneously strained Ge film and a film containing a misfit dislocation. One of these mechanisms which has not been reported in the literature is detailed.
The activation energy for this path is calculated to be 26\% smaller than the activation energy 
for half loop formation of a full, isolated 60$\degree$ dislocation. 
An extension of the common neighbor analysis method involving characterization of the geometrical arrangement 
of second nearest neighbors is used to identify and visualize the dislocations and stacking faults.    
\end{abstract}


\maketitle


\section{Introduction}

Transitions in complex systems typically involve multiple elementary steps between several intermediate states.
Each intermediate can be characterized as a local minimum on the energy surface of the system and each elementary
step can be characterized in terms of the minimum energy path connecting adjacent local minima along the transition path.  
Powerful optimization techniques have been developed for bringing an initial guess to the nearest minimum or nearest 
minimum energy path, but the challenge remains to find which set of local minima should be visited and in what order so as 
to form a transition path that involves the smallest increase in energy 
and, thereby, gives the lowest activation energy for the overall transition.

The formation and/or migration of extended defects in materials are prime examples of such complex transitions. 
A particularly important example is the formation of misfit dislocations in a strained Ge film deposited on a Si(001) substrate. This system can be used in a wide range of practical applications including photonic \cite{Hu2009,Chaisakul2014,Liu2015,Allred2014} and electronic \cite{Takagi2008} devices. 
It can also be used as an intermediate substrate for growing GaAs \cite{Kim1997,Carlin2000} or GaP \cite{Skibitzki2014} films on silicon.
The lattice constant of Ge is 4\% larger than that of Si so a large strain builds up in the Ge film preventing
coherent growth for more than a few monolayers (MLs).
The strain can be released by the formation of three-dimensional islands in the Stranski-Krastanow growth mode \cite{Eaglesham1990} 
which after coalescence gives a film with a high density of threading dislocations (TDs) 
(i.e. dislocations crossing the film) \cite{Sheldon1985} greatly deteriorating the film properties. 
Island formation can be prevented by low temperature deposition \cite{Eaglesham1991} 
or by the use of surfactants \cite{Bolkhovityanov2007}. 
The strain is then released by the formation of misfit dislocations (MDs). 
Some experiments have demonstrated the possibity of forming films with a low density of TDs \cite{Wietler2005,Myronov2007,Loh2007,Liu2012}, 
where the strain is released by a regular array of 90$\degree$ MDs 
(also known as edge or Lomer dislocations)
at the film-substrate interface \cite{Wietler2005,Liu2012}.
Such films can have high performance in various devices.
Since the 90$\degree$ MDs are sessile, i.e. they are not mobile, and should be placed at the Ge/Si interface 
while dislocations preferably nucleate at the surface, the formation mechanism is not direct.
An understanding of the formation mechanism could help optimize growth conditions of high performance films.

In the Ge/Si(001) system, mainly two types of MDs are efficient for relaxing the film strain.
They are characterized by the angle between the dislocation line and the Burgers vector. 
A 90$\degree$ MD is more efficient for releasing the film strain due to the fact that its Burgers vector lies in the (001) plane.  
A 60$\degree$ MD releases only half as much strain but it can glide on two of the 
(111)-type planes. 
A 60$\degree$ MD can be formed from the surface by the so-called half loop nucleation process.
Since 90$\degree$ MDs are sessile, they must form through the reaction of mobile dislocations.
Bolkhovityanov \textit{et al.} have presented a review of several mechanisms  proposed for the formation of 90$\degree$ MDs \cite{Bolkhovityanov2011}. In all these mechanisms, a 60$\degree$ MD forms first. Then, a complementary 60$\degree$ MD nucleates either independently 
or its nucleation is induced by the presence of the first MD \cite{Bolkhovityanov2013}. 
The two 60$\degree$ MDs then react together to form a 90$\degree$ MD. Such a reaction could, for example, 
be of the form:
\begin{equation}\label{eqReac}
 (a/2)[0\bar{1}\bar{1}](1\overline{1}1)+(a/2)[101](1\overline{1}\overline{1}) \rightarrow (a/2)[1\bar{1}0](001),
\end{equation}
where $a$ is the Si lattice constant, and $(a/2)[0\bar{1}\bar{1}](1\overline{1}1)$ indicates that the MD has a Burgers vector of $(a/2)[0\bar{1}\bar{1}]$ and that it glides on the $(1\overline{1}1)$ plane.

In early simulation work, the nucleation of a 90$\degree$ MD in a Ge/Si(001) was studied 
with a quasi-two-dimensional model \cite{Ichimuraa1995}.
Recently, a full three-dimensional simulation was carried out for a thin Ge film using the Stilling-Weber empirical potential
to describe the atomic interactions \cite{Trushin2015}.
By adding a repulsive bias potential \cite{Trushin2004,Trushin2009}, the relaxation of a coherent thin Ge/Si(001) film 
led to the formation of a straight 90$\degree$ MD at the Ge/Si interface. 
The closest minimum energy path (MEP) to this relaxation process was determined using the nudged elastic band 
(NEB) method \cite{Mills1995,Jonsson1998}. The transition path found in this way involves first the 
formation of a 60$\degree$ dislocation through the half loop nucleation process. 
Then, a complementary 60$\degree$ MD half loop nucleates and reacts with the first one to form the 90$\degree$ MD. 
The energy along the transition path shows two large scale barriers. The latter corresponds to the nucleation of the 
complementary MD and is 30\% lower than the first barrier. 
This mechanism thus corresponds to an induced nucleation of a complementary 60$\degree$ MD. 

Here, we revisit the problem of identifying the mechanism for the formation of a 90$\degree$ MD in the Ge/Si(001) system. 
Unlike a previous {\it local} optimization of the transition path \cite{Trushin2015}, 
we carry out a {\it global} optimization and generate a large number of intermediate candidate minimum energy structures. 
The procedure makes use of heredity transformations \cite{Oganov2006,Vilhelmsen2012}
where previously found atomic configurations are mixed to generate new, intermediate configurations, and calculations of
MEPs between the various configurations are carried out using a modified version of the NEB method. 
Several new mechanisms for the formation of 90$\degree$ MD are found in this way. Since this article focuses on the methodology, we describe only one of the mechanisms. The corresponding transition path for this mechanism involves one large scale energy barrier 
with an activation energy 26\% smaller than that of the half loop nucleation of an isolated 60$\degree$ MD. 
We also introduce a modified version of the common neighbor analysis (CNA) method \cite{Honeycutt1987,Clarke1993} and use it to identify dislocations and stacking faults in the diamond lattice. 
The modified CNA method is presented in Appendix A, and the modified NEB method in Appendix B.


\section{Methods}

\subsection{Model system}

A strained layer of Ge on a Si(001) substrate was modeled using the Stillinger-Weber potential function to 
describe the atomic interactions \cite{Stillinger1985}.
The parameters in the potential function were taken from Ref. \cite{Laradji1995}. 
This empirical potential function is, of course, only an approximation to the atomic interactions, but it is a commonly used approach for simulations of large Si and Ge systems. 
The system size needed to describe the phenomena studied here makes the computational effort for quantum mechanical density functional calculations prohibitively large.  We note that transition states and energy barriers tend to be overestimated by the 
Stillinger-Weber potential as it does not properly account for rehybridization \cite{Smith1995,Smith1996}. An overestimate of the activation energy for the dislocation nucleation can, therefore, be expected from calculations using the Stillinger-Weber potential. 

The system size considered here is $153.6 \times 153.6 \times 150$ \AA$^3$ and contains 80 000 atoms. Periodic boundary conditions are applied along the $x$ ($[1\bar{1}0]$) and $y$ ([110]) directions. 
The substrate contains 31 layers of Si in the $z$ ([001]) direction, and two bottom layers are kept fixed to represent
constraints due to lower layers that are not explicitly included in the simulation. 
We have checked that doubling the system size either in the $x$- or in the $y$-direction does not significantly change the energetics of the calculated transition paths. 
The coherent Ge film is 19 layers thick with a $p(2\times 1)$ dimer reconstruction of the surface. This film thickness corresponds to the critical thickness for the formation of a 60$\degree$ MD \cite{Trushin2015}.


\subsection{Global optimization procedure}

The procedure for optimizing the transition path is illustrated in Fig. 
1.
In order to explore the configuration space and identify possible 
intermediate configurations for the transition path, new minimum energy configurations are generated by heredity transformations
of known configurations. 
Two parent configurations are chosen from the set of previously determined minimum energy configurations. 
Initially, this set of configurations contains only the initial and the final states of the transition path.
A new configuration is then generated from the parent configurations by a heredity transformation which involves 
selecting coordinates of some of the atoms from one of the parent configurations 
(for instance by taking all the atoms which are outside a region defined by specifying two planes) 
while the coordinates of the rest of the atoms are selected from the other parent configuration. 
This procedure is analogous to approaches which have been developed for finding optimal structures of atomic
clusters \cite{Oganov2006,Vilhelmsen2012}.
The new configuration is likely to have high energy even after local minimization. However, when NEB calculations are 
carried out between the parent and child configurations, dips in the energy curve point to additional and possibly lower
energy minima. The intermediate configurations that are found to have low energy become included in the set of possible 
intermediate configurations along the transition path, and can be used as parent configurations in subsequent 
heredity transformations.

NEB calculations of MEPs are carried out between low energy intermediate configurations that have been identified. 
Again, each NEB calculation can reveal new
low energy configurations that are added to the set of possible intermediate configurations in the optimal path.

The choice of the cutting planes for the heredity transformation was made intuitively keeping in mind that 
the goal is to generate and select configurations which
can be visited along the path from the initial to the final configuration. The best results were obtained using planes from the (111)-type family.
It would be of great interest to find an efficient way to automate these 
choices as is sometimes done in genetic algorithms. 
The challenge here is that the algorithm would need to handle both a set of configurations and 
a set of paths and would need to use an objective function taking both the energy and the connectivity of the configurations into account.

\begin{figure}
\includegraphics[width=0.8\linewidth]{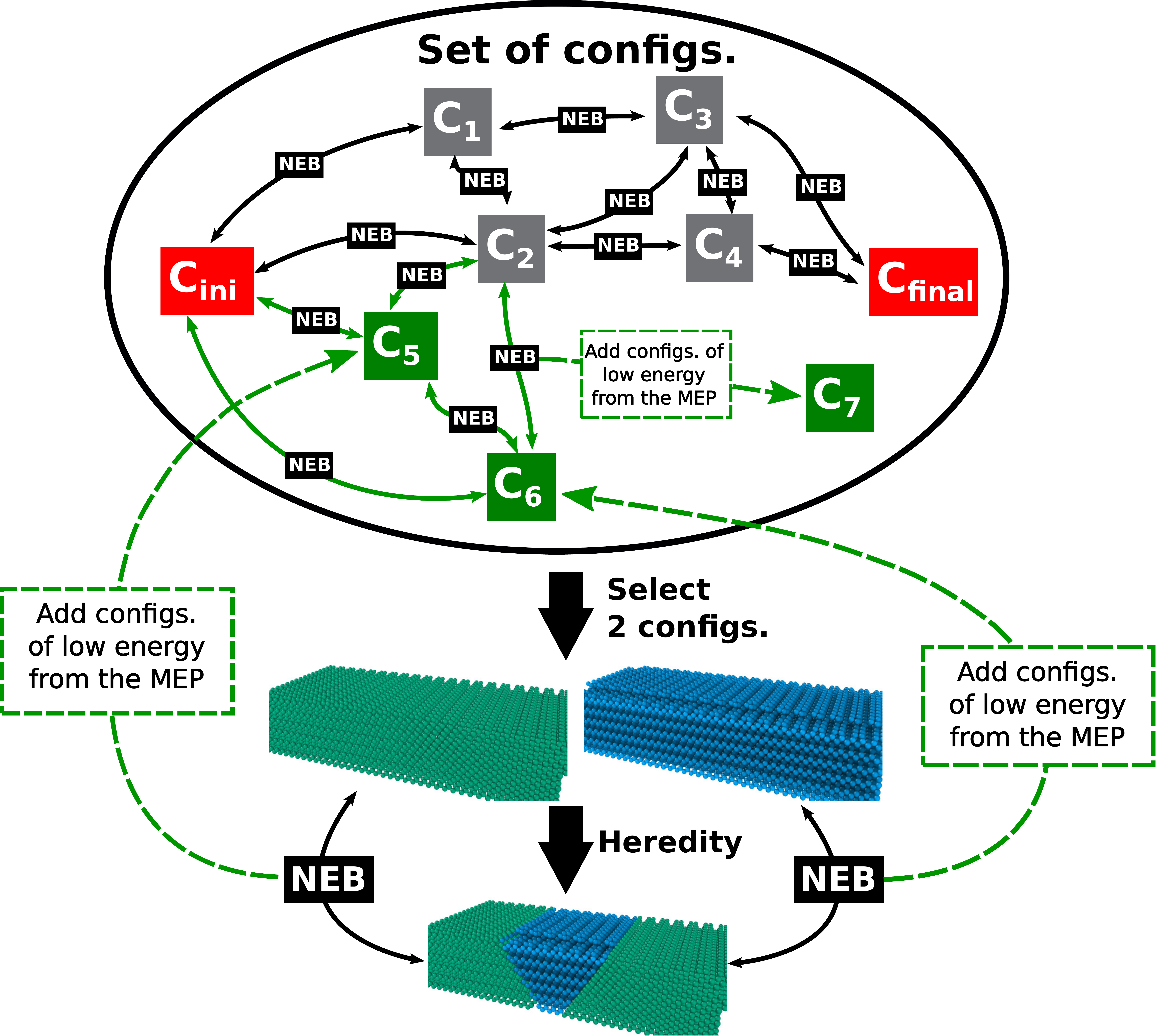}
\label{fig:Method}
\caption{
Illustration of the procedure used for the global optimization of the transition path between the initial (C$_{\rm ini}$) and final (C$_{\rm final}$) configurations. The procedure is based on generating a set of low energy intermediate configurations (C$_i$) for the path. In the end, the
optimal transition path is constructed from a  
sequence of minimum energy paths between a subset of the intermediate configurations so as to 
give the smallest rise in energy along the path and, thereby, the smallest activation energy for the transition.
In this example, the set initially contains 4 intermediate configurations, {C$_1$, C$_2$, C$_3$ and C$_4$}. In order to generate new intermediate configurations, two parent configurations are selected from the set and a heredity transformation is carried out to generate a new, child configuration. 
After local minimization, minimum energy paths are calculated between each parent and child 
configuration and new low energy configurations C$_5$ and C$_6$ appearing as minima along the paths are added to the set. Additional low energy configurations can also be found from 
NEB calculations of
minimum energy paths between pairs of low energy configurations. In the present example C$_7$ is extracted from an NEB calculation between C$_2$ and C$_6$. 
By adding more and more intermediate configurations, the potential energy surface is further explored and improved path are progressively found.}
\end{figure} 


The procedure can be best illustrated by considering the early stage of the global optimization. We consider a transition between an initial commensurate configuration shown in Fig. 2a  and a final configuration with a straight 90\degree MD crossing the system shown in Fig. 2b . In order to visualize the dislocation, the second neighbor CNA method described in Appendix A
is used to identify atoms that are locally not in a perfect diamond lattice environment. Only these atoms are then rendered.
 A NEB calculation starting from a linearly interpolated path between these two configurations leads to a path in which an overextended dislocation half-loop forms. Because of this overspreading, the activation energy is larger than 50 eV and depends on the system size. The nucleation event can be localized using the procedure described in Fig. 1. A configuration is generated through a heredity transformation by selecting coordinates from the final configuration for all the atoms which are in between the two green planes shown in Fig. 3a, and from the initial commensurate configuration for the other atoms. The resulting configuration after relaxation is shown in Fig. 3a. This configuration features a short segment of a 90$\degree$ MD terminated by two threading dislocations. It has an energy almost 120 eV larger than the commensurate configuration. However, when running an NEB between the initial configuration and this configuration, we find along the path a configuration having a lower energy than the initial configuration. This configuration is shown in Fig. 3b. It features defects (dislocations and a stacking fault) which constitute a localized seed for the formation of a straight 90\degree MD crossing the whole system.
By using different planes in the heredity transformation, several nonequivalent intermediate configurations of low energy can be found an used to optimize the transition path.

\begin{figure*}
\includegraphics[width=0.9\textwidth]{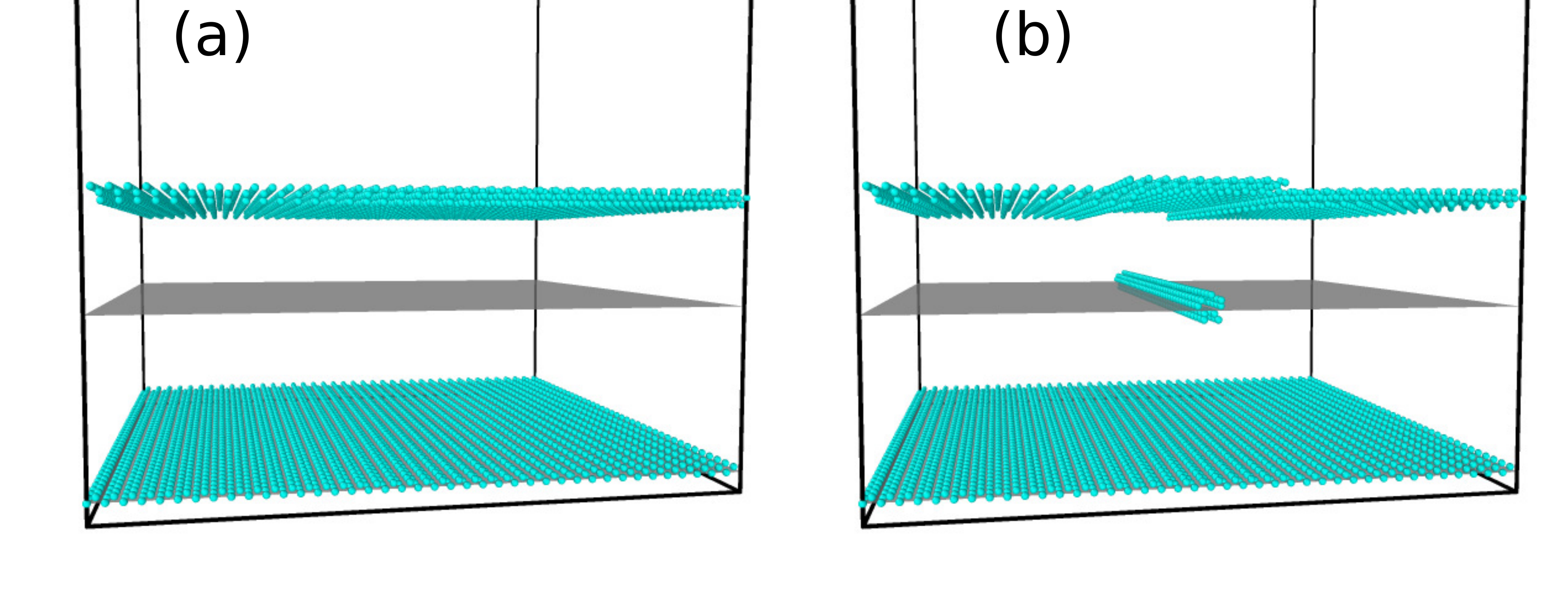}
\label{fig:90Nucl} 
\caption{
Initial commensurate configuration (a) and final configuration with a straight 90$\degree$ MD crossing the system.
Only atoms in a local environment that does not 
correspond to a diamond lattice are represented.  The dark gray (001) plane corresponds to the substrate-film interface. }
\end{figure*}

\begin{figure*}
\includegraphics[width=0.9\textwidth]{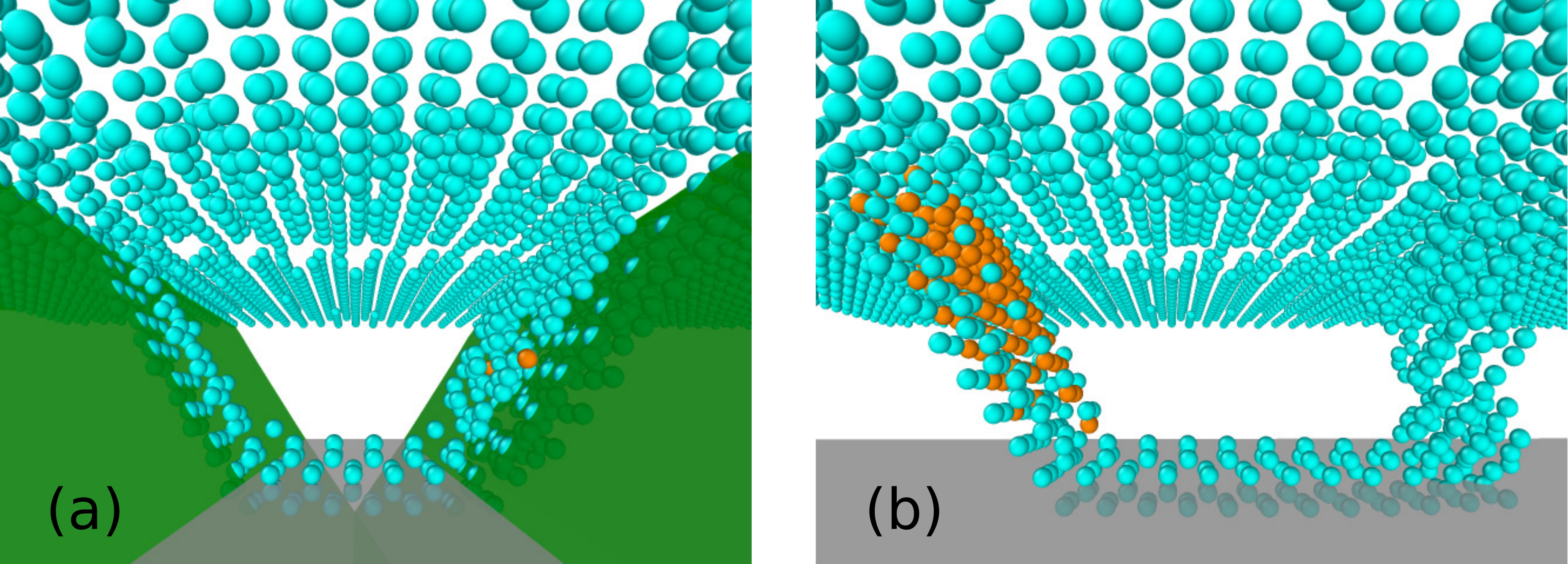}
\label{fig:90Nucl} 
\caption{
Configuration generated by heredity transformation (a). Along the path obtained from a NEB calculation between this configuration and the initial defect free configuration (Fig. 2a), a configuration of low energy is found (b).
Only atoms in a local environment that does not 
correspond to a diamond lattice are represented. The orange atoms are those whose local environment corresponds to a hexagonal diamond lattice. The green planes, are the planes which were used in the heredity transformation. The dark gray (001) plane corresponds to the substrate-film interface. }
\end{figure*}

\subsection{Results}

An optimization of the transition path for the formation of a straight 90$\degree$ dislocation at the interface between the Ge film and the Si substrate was carried out and several mechanisms were identified from the calculations.
Since this article focuses on the methodology, only one mechanism which has not been reported previously is shown.  Its energy profile is presented in Fig. 
4a with a close-up on the first part of the path in Fig. 4b.
The calculated formation mechanism corresponding to this path has an activation energy of $39.8$ eV. 
\begin{figure}
\includegraphics[width=\linewidth]{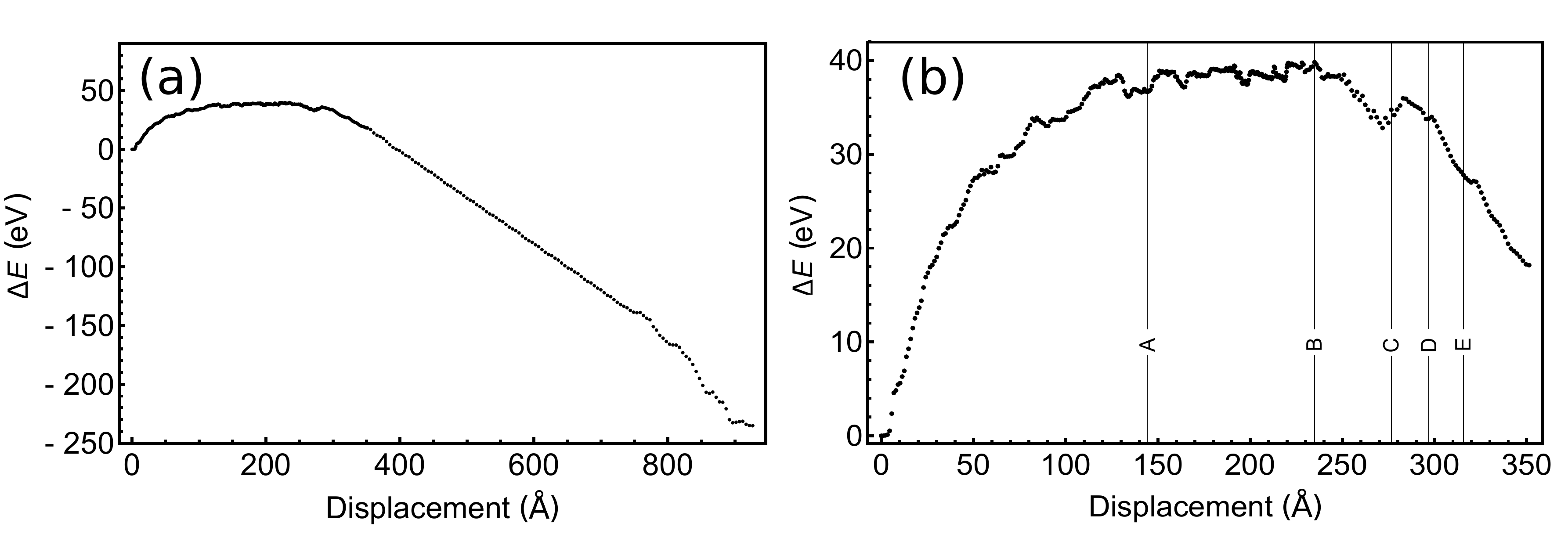}
\label{fig:NRJProfMyPath}
\caption{
Energy along the optimal transition path found for the formation of a 90$\degree$ MD as a function of the cumulative displacements of the atoms. 
A total of 504 intermediate configurations represent the minimum energy path obtained 
from the NEB calculations. Panel (b) is a close-up of panel (a).
The vertical lines in panel (b) indicate the configurations represented in Figs. 
5 and 6. 
}
\end{figure} 

Some configurations along the transition path are represented in Fig. 
5.
The dislocation extraction algorithm (DXA) \cite{Stukowski2010b,Stukowski2012}
is used to determine the dislocations' Burgers vectors.
A schematic representation of the configurations is provided in Fig. 
6
and an animation of the MEP is provided in the Supplementary Information.
The mechanism starts with the nucleation of a dislocation half loop with Burgers vector $(a/2)[0\bar{1}\bar{1}]$ corresponding to that of a 60$\degree$ MD. 
The half loop forms by a glide in the $(1\bar{1}1)$ plane but one of the threading arms then 
undergoes a cross-slip in the $(11\bar{1})$ plane 
(see configuration A in Figs. 
5 and 6). The energy of this configuration is 36.7 eV higher than the initial state, 
close to the maximum energy of the optimal transition path. 
Then, the dislocation reacts to form a Shockley partial dislocation and a double Shockley partial dislocation according to the following reaction:
\begin{equation}
 (a/2)[0\bar{1}\bar{1}]\rightarrow (a/6)[2\bar{1}1] +(a/3)[\bar{1}\bar{1}\bar{2}].
  \label{eq:90ReacFirstPart}
\end{equation}
This reaction takes place in the $(11\bar{1})$ plane which is a slip plane for all these dislocations.
It is initiated from the surface and results in the formation of a stacking fault as can be seen in configuration B in Fig. 
5.\footnote{The stacking fault energy calculated with the Stillinger-Weber potential is zero. 
The experimentally measured intrinsic stacking fault energy varies from about 3.7 to 6.3 meV/\AA$^2$ \cite{George1987}. 
By adding this stacking fault energy to our model, the activation energy would increase by less than 2 eV.}
The stacking fault can be identified easily with the CNA method since the local environment of atoms within the stacking 
fault corresponds to a hexagonal diamond lattice. Configuration B, shown in Figs. 
5b ad 6b, has the highest energy along the transition path, 39.8 eV. 
Then both partial dislocations glide in the $(11\bar{1})$ plane until the single Shockley and the double Shockley partial dislocations are oriented in the [011] and [101] directions, respectively. The stacking fault then spans over all the $(11\bar{1})$ surface between the [001] and [101] directions (see configuration C in Figs.
5c, 5d and 6c]. 

\begin{figure*}
\includegraphics[width=0.9\textwidth]{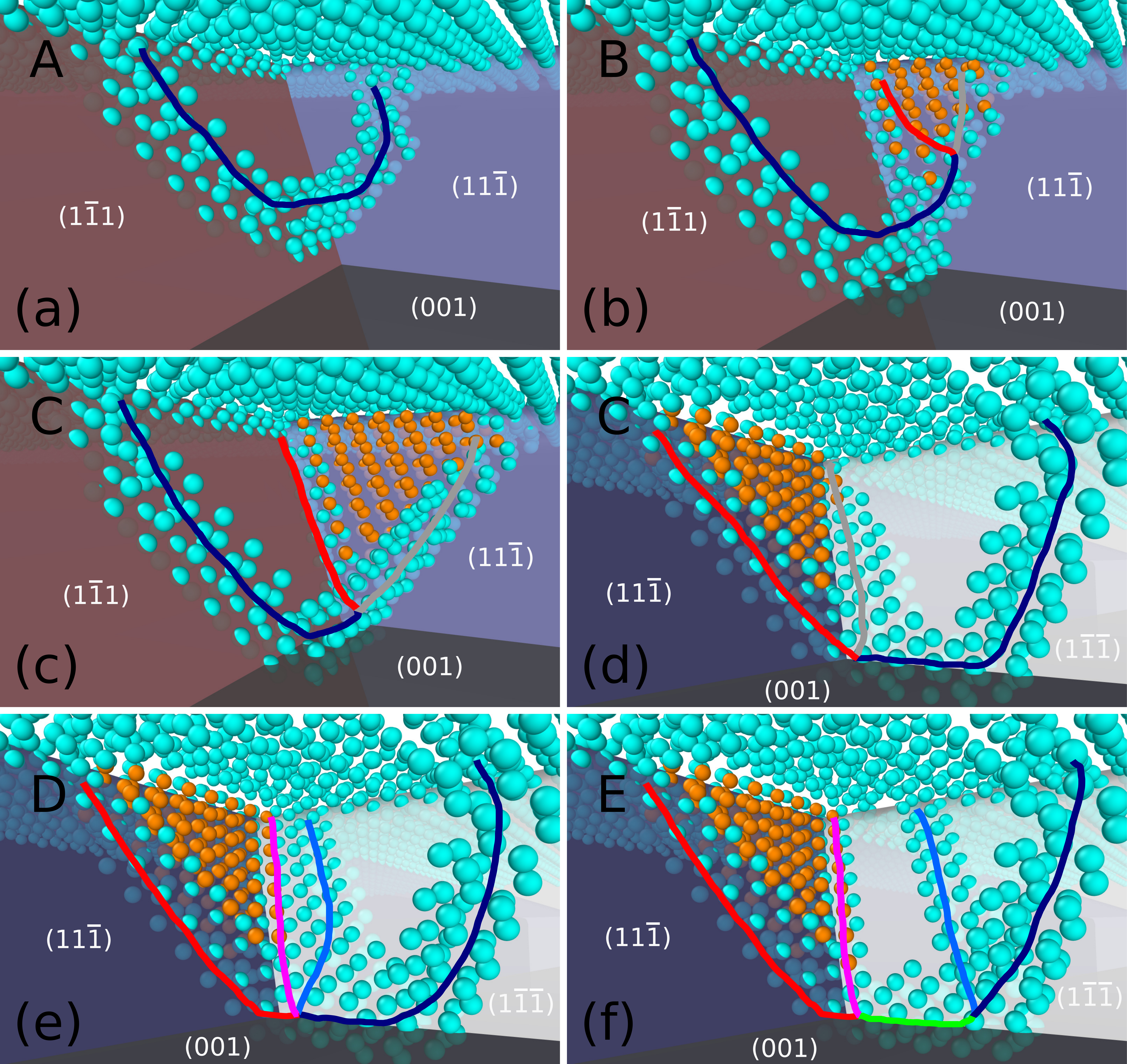}
\label{fig:90Nucl} 
\caption{
Intermediate configurations along the optimal path found for the formation of a 90$\degree$ dislocation in Ge on Si(001). 
Only atoms in a local environment that does not 
correspond to a diamond lattice are represented. The orange atoms are those whose local environment corresponds to a hexagonal diamond lattice. The dark blue and light blue lines indicate dislocation lines with a $(a/2)[0\bar{1}\bar{1}]$ and a  $(a/2)[\bar{1}0\bar{1}]$ Burgers vector corresponding to that of a 60$\degree$ MD. The red, gray, magenta and green lines indicate a Shockley partial, a double Shockley partial, a Shockley partial and a 90$\degree$ dislocations, respectively.
The red, blue and white planes correspond to the $(1\bar{1}1)$, the $(11\bar{1})$ and the $(1\bar{1}\bar{1})$ planes, respectively. The dark gray (001) plane corresponds to the substrate-film interface. 
The energy of each configuration is indicated by the vertical lines in Fig. 
4b.
Configuration B  is the highest energy configuration along the optimal transition path. 
Panels (c) and (d) show two different views of configuration C.
}
\end{figure*}

\begin{figure*}
\includegraphics[width=0.5\textwidth]{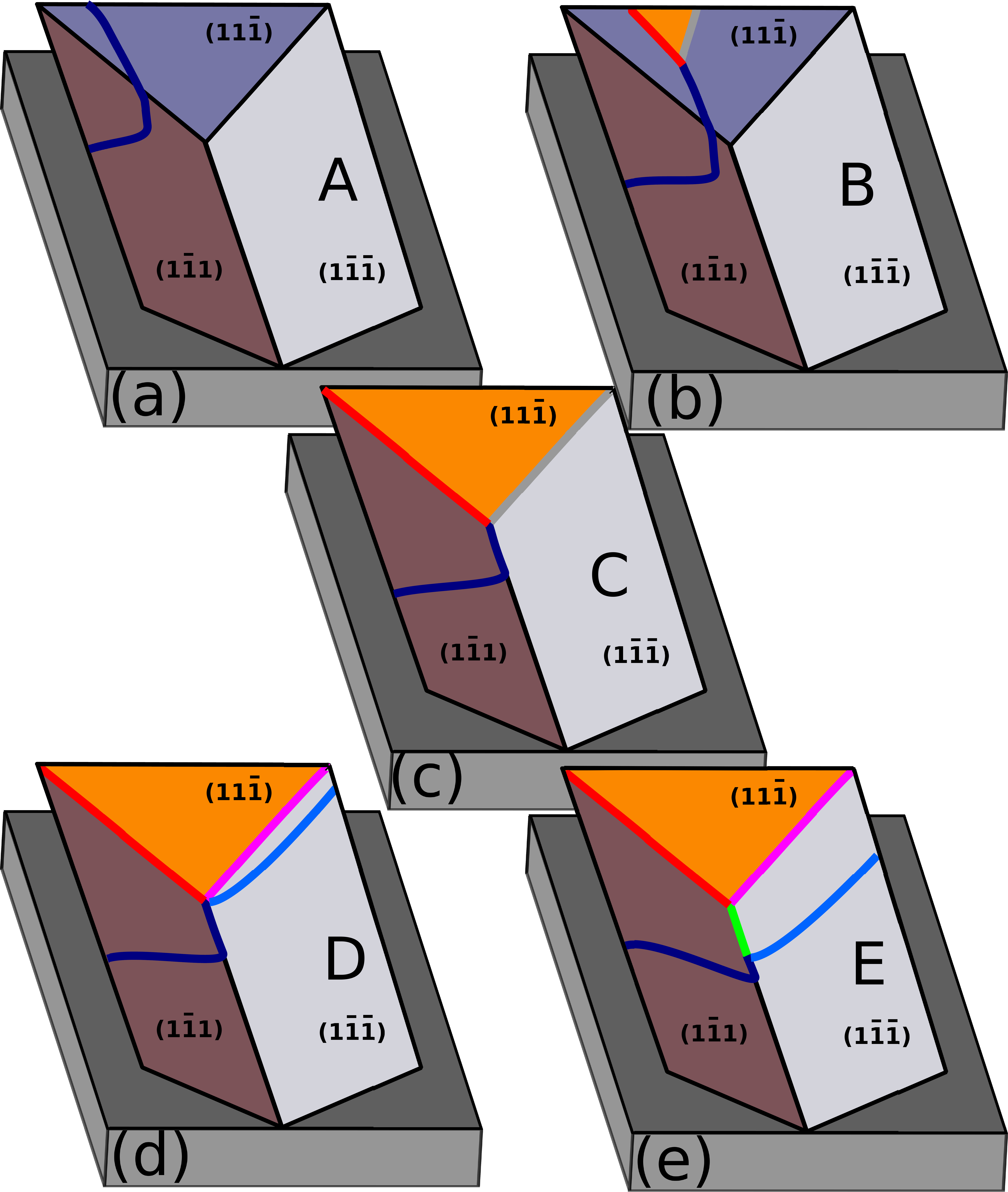} 
\label{fig:90NuclSchem}
\caption{ 
Schematic view of the intermediate configurations along the optimal path 
found for the formation of a 90$\degree$ dislocation in Ge on Si(001). 
The dark blue and light blue lines indicate dislocation lines with a $(a/2)[0\bar{1}\bar{1}]$ and a  $(a/2)[\bar{1}0\bar{1}]$ Burgers vector corresponding to that of a 60$\degree$ MD. The red, gray, magenta and green lines indicate a Shockley partial, a double Shockley partial, a Shockley partial and a 90$\degree$ dislocations, respectively.
The red, blue and white planes correspond to the $(1\bar{1}1)$, the $(11\bar{1})$ and the $(1\bar{1}\bar{1})$ planes, respectively. The gray plane corresponds to the substrate-film interface. 
The energy of each configuration is indicated by a vertical line in Fig. 
4b.
}
\end{figure*}

The double Shockley partial dislocation then reacts to form a Shockley partial and a dislocation with a $(a/2)[\bar{1}0\bar{1}]$ Burgers vector corresponding to that of a 60$\degree$ MD
(from configuration C in Figs. 
5c, 5d and 6c 
to configuration D in Figs. 
5e and 6d) following
\begin{equation}
 (a/3)[\bar{1}\bar{1}\bar{2}] \rightarrow  (a/6)[1\bar{2}\bar{1}] + (a/2)[\bar{1}0\bar{1}].
  \label{eq:90ReacSecondPart}
\end{equation}
By combining Eqs. (\ref{eq:90ReacFirstPart}) and (\ref{eq:90ReacSecondPart}) and reversing 
the direction for one dislocation which changes its Burgers vector from $(a/2)[\bar{1}0\bar{1}]$ to $(a/2)[101]$, 
we have
\begin{equation}
 (a/2)[0\bar{1}\bar{1}]+(a/2)[101] \rightarrow (a/6)[2\bar{1}1]+(a/6)[1\bar{2}\bar{1}],
  \label{eq:90ReacSecondPart}
\end{equation}
which describes the dislocation node in configuration D (see Figs. 5e and 6d).
Then, by gliding of the threading dislocation on the $(1\bar{1}\bar{1})$ plane, the 60$\degree$ MD reacts with the other 60$\degree$ MD 
to form the 90$\degree$ MD (from configuration D in 
Figs. 5e and 6d 
to configuration E in 
Figs. 5f and 6e) according to Eq. (\ref{eqReac}).

The 90$\degree$ MD can then grow by glide of both 60$\degree$ threading dislocations on their respective glide planes. The growth of the 90$\degree$ MD causes a linear decrease of the energy along the transition path which can be seen in Fig. 4a for distances from 300 to 750 \AA.

Finally, due to the periodic boundary conditions used in our system, the gliding threading dislocations meet and react with the partial dislocations until only a straight 90$\degree$ MD remains at the interface. This reaction causes a large decrease in energy due to the annihilation of the threading dislocations. This is not described in detail here since it is a consequence of the periodic boundary conditions. 


\section{Discussion}



The energy barrier obtained here, 39.8 eV, is too high for it to be overcome by thermal fluctuations at experimentally relevant temperatures.
However, this does not mean that the corresponding mechanism is irrelevant. Many atomistic calculations of the nucleation of dislocations in semi-conductor materials predict high activation energies \cite{Ichimuraa1995,Bolkhovityanov2001,Hull1989,Li2010}.

One possible explanation for the large barriers is that dislocations form from existing defects whereas our calculations were carried out starting from a defect-free system. 
The activation energy could be significantly lowered by including defects in the initial configuration. When starting from a defect free film (Fig. 2a) as was done in this study, the formation of a straight 90$\degree$ MD leads to the formation of double-layer steps on the surface as shown in Fig. 2b. This is energetically unfavorable since step atoms have dangling bonds. However, if the initial film contains a vacancy stripe, the formation of the 90$\degree$ MD can eliminate the surface steps and new bonds are formed. Furthermore, steps act as stress concentrators. \cite{Brochard2010,Godet2004,Godet2006,Godet2009,Shima2010,Li2010,Li2012,Izumi2008,Marzegalli2005,Marzegalli2005a}. \begin{figure*}
\includegraphics[width=0.5\textwidth]{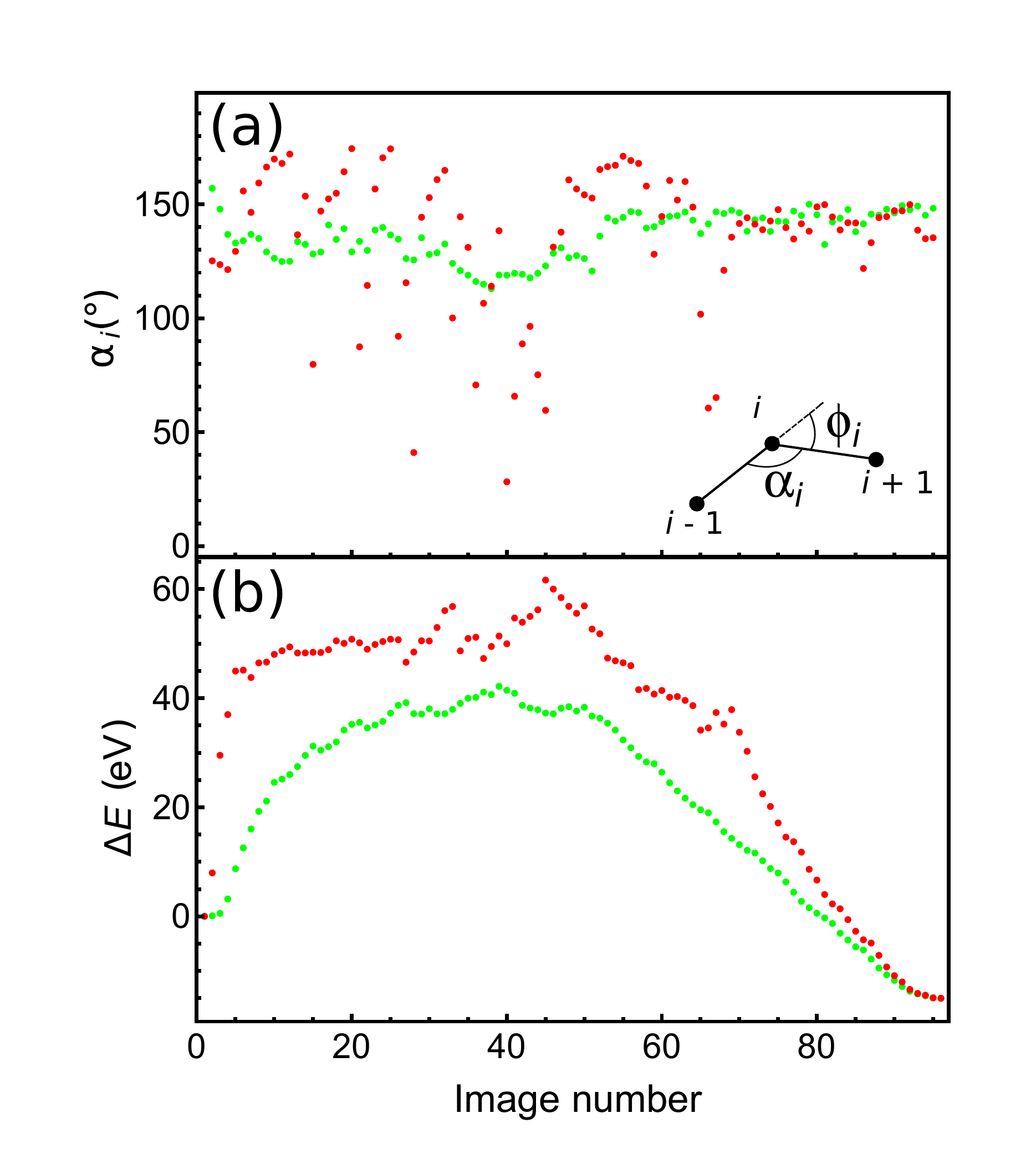} 
\label{fig:InfluenceOfStep}
\caption{Side view of four configurations for a Ge/Si(001) film. Starting from  a film with surface steps (a), the formation of a straight 90$\degree$ MD leads to the elimination of the steps (b) and thereby, high energy dangling bonds. Starting from a film containing a vacancy stripe (c), the formation of a straight 90$\degree$ MD leads to the disappearance of the vacancy stripe (d).}
\end{figure*}
Preliminary calculations indicate that the presence of steps on the surface lowers the activation energy by about 12 eV. Points defects such as impurities can also facilitate the nucleation of dislocations \cite{Barnoush2010}.

Another explanation for the large activation energy is the inaccuracy of the empirical potential. To the best of our knowledge, the only empirical potentials which have been parameterized for GeSi are the SW and the Tersoff potentials \cite{Tersoff1988,Tersoff1989,Tersoff1990}. The issue is that no dislocations were included in the training set for these parameterizations. These potentials should therefore not be expected to properly describe rehybridization at the dislocation core and they are expected to overestimate the dislocation core energy. Preliminary calculations using the Tersoff potential show that the activation energy is lowered by about 10 eV as compared to the SW potential. When comparing different mechanisms, the difference in activation energy is found to be similar with the SW and Tersoff potentials. Despite the lack of quantitative accuracy of the semi-empirical potentials, we expect that the main features of the mechanisms found for dislocation formation could still hold.

The mechanisms for the formation of 90$\degree$ MDs previously presented in the literature 
require as a first step the formation of a 60$\degree$ MD \cite{Bolkhovityanov2011} which is usually assumed to 
occur through the half loop nucleation process. 
In a previous study of the same system and using the Stillinger-Weber potential,
it was shown that the half loop nucleation of a 60$\degree$ MD has an activation energy of 54 eV \cite{Trushin2015}. 
Since the new mechanism presented here has a significantly lower activation energy, 39.8 eV, we 
can conclude that unless a more favorable mechanism for the formation of a 60$\degree$ MD exists, 
the formation of a 90$\degree$ MD through the mechanism presented in Figs. 
5 and 6
is more likely than the formation of a 60$\degree$ MD. 
Furthermore, unlike previously reported mechanisms, the mechanism identified here requires the 
crossing of only one large scale energy barrier.
This mechanism could thus explain the fact that for Germanium rich GeSi films on Si(001) substrates 
mainly 90$\degree$ MDs and only a few 60$\degree$ MDs have been experimentally observed \cite{Myronov2007,Liu2012,Marzegalli2013}. 

When 90$\degree$ MDs form by this new mechanism, a stacking fault should be left in the film. 
Experimentally stacking faults are, however, rarely found for Ge/Si(001) films and when they are observed \cite{Yamasaki2004,Bharathan2013,Legoues1990,Hiroyama1998,Kim1996}, 
they are assumed to form either because of the presence of impurities \cite{Bharathan2013,Legoues1990}, 
or because of island formation in the Stranski-Krastanow growth mode \cite{Hiroyama1998}, 
or by the presence of an amorphous layer \cite{Kim1996}.  
A possible explanation for the fact that stacking faults are not observed experimentally 
is that they disappear after the formation of the 90$\degree$ MD. 
Indeed, two Shockley partial dislocations can glide toward each other on the $(11\bar{1})$ plane and react to form a 90$\degree$ dislocation. 
A second explanation could be that experimentally, a low temperature buffer layer is often used 
prior to the high temperature deposition of the complete film \cite{Myronov2007,Loh2007,Liu2012,Bharathan2013}. 
Bolkhovityanov \textit{et al.} have suggested that the buffer layer 
already contains 60$\degree$ MDs which initiate the formation of 90$\degree$ MDs through the induced nucleation of 
complementary 60$\degree$ MDs \cite{Bolkhovityanov2012a}.
We expect that the mechanism presented here
should be dominant when the film is very thin, i.e. at early stages of Ge film growth.
Assuming that the shape of the stacking fault is independent of the film thickness, $h$, 
the area of the stacking fault and the stacking fault energy are proportional to $h^2$. 
Clearly, there will be a critical 
thickness above which the mechanism presented here becomes less efficient than other known mechanisms.


\section{Summary}

We have presented an approach for the global optimization of a transition paths in complex systems. It involves generating 
a number of possible intermediate configurations using heredity transformations of pairs of known minimum energy
configurations. Minimum energy paths are calculated using a revised NEB method and low energy intermediate configurations identified from dips in the energy along the paths. The revised NEB includes part of the spring force 
to prevent the path from forming acute angles between adjacent segments and thereby improves the convergence for long and complex transition paths, where the density of discretization images is low compared to variations in the path tangent.

The application of this approach to the formation of a 90$\degree$ dislocation in a heteroepitaxial Ge/Si(001) film has
revealed a mechanism unlike those reported previously. After the nucleation of a small dislocation half loop with a Burgers vector corresponding to that of a 60$\degree$ MD, 
one end of the loop splits into a Shockley and a double-Shockley partial. The double Shockley partial dislocation later on splits into one Shockley partial and a  dislocation  with a Burgers vector corresponding to that of a complementary 60$\degree$ MD which then reacts with the other threading arm of the initial dislocation to form a 90$\degree$ MD.
One characteristic feature of this mechanism is that it induces the formation of a stacking fault. 
For a 19 ML thick film, this mechanism has an activation energy which is 
26\% smaller than a half loop nucleation of an isolated 60$\degree$ MD.

Finally, an extension of the CNA method for systematically characterizing the local environment of atoms in a diamond lattice has been presented and used here to identify and visualize MDs and stacking faults. The method is based on the characterization of the geometric 
arrangement of second nearest neighbors and has been implemented in the OVITO software \cite{Stukowski2010b}.


\section{Acknowledgments}
This work has been supported in part by the Academy of Finland through its COMP CoE (T.A-N., no. 251748 and 284621) and FiDiPro (E.M. and H.J., no. 263294) grants. O.T. was supported by the Russian Foundation for Basic Reserch grant No. 14-00139a.
We acknowledge computational resources provided by the Aalto Science-IT project and CSC IT Center for Science Ltd in Espoo, Finland.


\appendix

\section{Common Neighbor Analysis}
\label{app:CNA}

To identify and visualize the dislocations and stacking faults, we use the CNA method applied to second nearest 
neighbors. The CNA involves classification of pairs of atoms according to their local environment. Three indices
characterize each pair.  The first two give the number of common neighbors and the number of bonds 
(i.e. atom distance below a given cutoff distance) formed between the common neighbors \cite{Honeycutt1987} and the
third index is the longest continuous chain of common neighbor bonds \cite{Clarke1993}. When applied to systems
with close packing, such as materials that crystallize in FCC, HCP and BCC structures, the neighbors used in the 
analysis are nearest neighbors. 
For example, an atom locally in an FCC environment forms 12 pairs of the 421 type with its nearest neighbors. 

However, this method is not directly suitable for materials crystallizing in the diamond 
structure, because nearest neighbor pairs of atoms do not possess any common neighbors. 
Even though it is possible to work around this problem by extending the cut-off range and taking into account second nearest neighbors \cite{Stukowski2012a}, the second and third neighbor shells in the diamond lattice are not well separated. 
Hence, even a small elastic strain or thermal displacements easily disturb the computed CNA fingerprints, 
rendering the method unreliable. 
The extended CNA method used here exploits the fact that the cubic diamond structure consists of two interleaved FCC lattices. 
The analysis algorithm is as follows: 
Given an atom whose local environment is to be classified, 
the nearest neighbors are first identified. An atom in a perfect diamond lattice has four nearest neighbors. 
For each neighbor, its nearest neighbors are in turn identified. 
In the diamond lattice, this gives three additional atoms, excluding the central atom whose environment is being characterized, 
so there are a total of 12 second nearest neighbors. In a
perfect diamond lattice, these 12 second nearest neighbors all form 421 pairs with the central atom. 
The nearest neighbors are not included in this analysis. 
The cutoff radius is taken to be $r_{\rm CNA}= \hat{r}(1+\sqrt{2})/2$, where $\hat{r}$ is the average distance of 
the 12 second nearest neighbors from the central atom \cite{Stukowski2012a}. 

If the FCC signature is detected, the central atom is marked as being in a local environment
belonging to a perfect cubic diamond lattice. If the 12 atom pairs correspond to the HCP lattice, 
the central atom is marked as an atom in a hexagonal diamond lattice. 
This method for the identification of atoms in a diamond lattice is computationally efficient, 
has no adjustable parameters, and is insensitive to small perturbations in the atomic positions. 
Since second neighbors are taken into account by the method, it is possible to discriminate between cubic and 
hexagonal diamond arrangements, making it possible to identify stacking faults.  

This method has been implemented in the visualization software OVITO \cite{Stukowski2010b} and can be used in the
dislocation extraction algorithm \cite{Stukowski2010b,Stukowski2012} to automatically identify dislocation defects 
and their Burgers vectors in crystals with diamond structure.


\section{Modified NEB method}
\label{app:modNEB}

The NEB method \cite{Mills1995,Jonsson1998,Jonsson2011} can be used to relax a given initial path to the nearest MEP. 
The path is discretized by creating a set of $N$ configurations (or 'images') of the system. Each image, $i$, 
is defined by its position vector $\bm{R}_i$. 
An intermediate image (i.e. $1<i<N$) interacts with its previous and subsequent image by an artificial spring
to control the distribution of images along the path.  
The component of the gradient of the atomic interaction energy, $V(\bm{R}_i)$, 
along the tangent is removed to prevent it from interfering with the way the springs distribute the images. 
In the usual implementation of the NEB,
the spring force on image $i$ is applied only along the tangent direction of the path $\bm{\hat{\tau}}_i$ 
in order to prevent the chain from cutting corners where the MEP is curved. 
The total force acting on an intermediate image is then given by
\begin{equation}
 \bm{F}_i = -  \bm{\nabla} V(\bm{R}_i)\vert_{\bot }+\bm{F}_i^s\vert_{\parallel },
  \label{eq:OriginalNEB}
\end{equation}
where the perpendicular component of the gradient is given by
\begin{equation}
 \bm{\nabla } V(\bm{R}_i)\vert_{\bot } = \bm{\nabla } V(\bm{R}_i) -  \bm{\nabla } V(\bm{R}_i) \cdot \bm{\hat{\tau}}_i \bm{\hat{\tau}}_i,
  \label{eq:Grad}
\end{equation}
and the spring force is calculated as
\begin{equation}
  \bm{F}_i^s\vert_{\parallel }= k\left( \left| \bm{R}_{i+1}-\bm{R}_i\right|- \left| \bm{R}_{i}-\bm{R}_{i-1}\right| \right)\bm{\hat{\tau}}_i,
  \label{eq:SpringForce}
\end{equation}
with $k$ being the spring constant.

Provided the number of images is large enough, the relaxation of the chain converges to an MEP. 
However, when a transition path involves a large number of intermediate minima and the number of images is small compared to the complexity of the path, the convergence of the NEB method can be problematic.  
Fig. B.8
shows the progress of an NEB calculation of a path for the 90$\degree$ misfit dislocation formation in Ge/Si(001) 
in terms of the maximum force acting on an image, $\max\left|\bm{F}_i\right|$ 
(i.e. the maximum value over all images of the norm of the atomic force vector). 
The NEB calculation has not converged after 30 000 steps since the maximum force acting on an image along the reaction path 
remains large, even up to 10 eV/\AA. 
This lack of convergence results from the fact that the density of images is low compared to the local
curvature of the MEP and the angle $\alpha_i$ formed between adjacent segments of the path (see inset in Fig.B.9a 
)
can be acute. 
The estimate of the path tangent is then inaccurate and unreliable results can be obtained from the NEB calculation.

\begin{figure}
\includegraphics[width=0.8\linewidth]{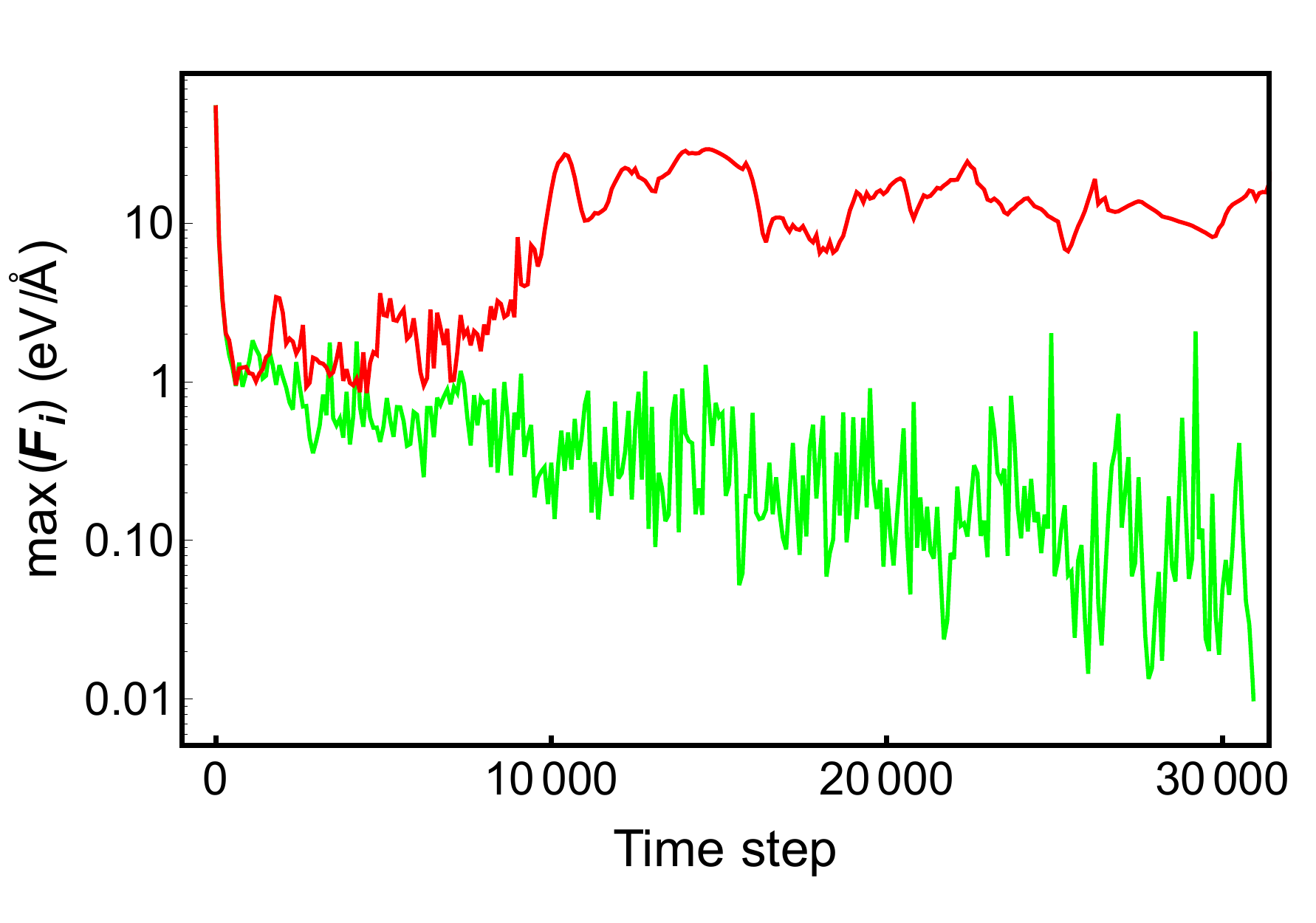} 
\label{fig:Convergence}
\caption{ 
The magnitude of the maximum of the force acting on an image in an NEB calculation as a function of the 
number of velocity projection optimization \cite{Jonsson1998} iterations. 
Results from a regular NEB calculation \cite{Henkelman2000a} (red) and from the modified NEB (green) are shown for comparison.
}
\end{figure} 

\begin{figure}
\includegraphics[width=0.9\linewidth]{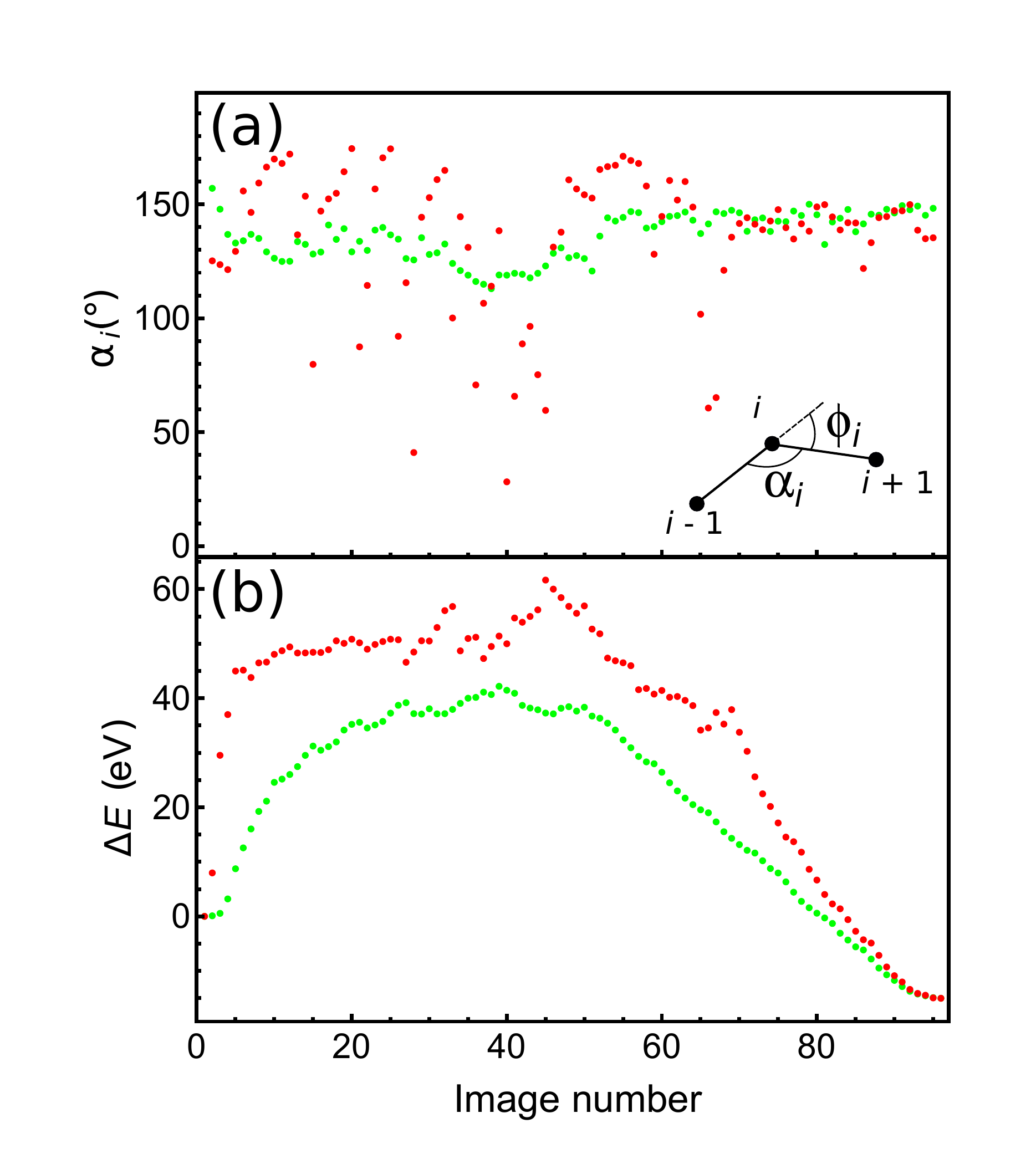} 
\label{fig:PathAngle} 
\caption{
(a) The angle between adjacent segments of the NEB path at the various images, $i$, 
after 30 000 steps of NEB relaxation. The definition of the angle is shown in the inset. 
Results from a regular NEB \cite{Henkelman2000a} (red) and from the modified NEB (green) are shown.
(b) Energy along paths obtained after 30 000 iterations of an NEB minimization. 
Results from a regular NEB \cite{Henkelman2000a} (red) and from the modified NEB (green) are shown.
}
\end{figure} 

We thus modify the NEB to maintain the path straight enough by applying part of the spring force in directions normal to the path as has been suggested earlier \cite{Jonsson1998}.  However, the present implementation differs from the 
previous one in that the improved tangent definition is used \cite{Henkelman2000a}, 
as it has been shown to significantly improve the convergence of NEB calculations. 
The force acting on an image is then given by
\begin{equation}
 \bm{F}_i = -  \bm{\nabla } V(\bm{R}_i)\vert_{\bot}+ \bm{F}_i^s\vert_{\parallel }+f(\phi_i) \bm{F}_i^{s_2}\vert_{\bot} ,
  \label{eq:NewNEB}
\end{equation}
where $\bm{F}_i^{s_2}\vert_{\bot}=\bm{F}_i^{s_2}-\bm{F}_i^{s_2} \cdot \bm{\hat{\tau}}_i \bm{\hat{\tau}}_i$ 
is the perpendicular component of the spring force
\begin{equation}
 \bm{F}_i^{s_2} = k \left[ \left( \bm{R}_{i+1}- \bm{R}_{i} \right)- \left(\bm{R}_{i}-\bm{R}_{i-1}  \right)\right], 
 \label{eq:SpringForceNew}
\end{equation}
and $ f(\phi_i)$ is a switching function which goes linearly from zero if the path is straight to unity if adjacent
segments of the path form a right angle. It is defined by
\begin{equation}
  f(\phi_i) = {1 \over 2} \left(1+\cos\left(\pi \cos(\phi_i)\right) \right),
  \label{eq:AngularContr}
\end{equation}
when $0<\phi_i<\pi/2$, and $ f(\phi_i) = 1$ for $\phi$>$\pi$/2. The $\cos(\phi_i)$ characterizes the path angle:
\begin{equation}
\cos(\phi_i)=\left( \bm{R}_{i+1}- \bm{R}_{i} \right) \cdot \left(\bm{R}_{i}-\bm{R}_{i-1}  \right)
/\left(\left|\bm{R}_{i+1}- \bm{R}_{i} \right| \left|\bm{R}_{i}-\bm{R}_{i-1}  \right|\right).
\end{equation}
If the path is straight, $f(\phi_i)=0$ and the force acting on an image is then equivalent to Eq. (\ref{eq:OriginalNEB}). 
Otherwise a perpendicular contribution from the spring force is taken into account. 
As illustrated in Fig. B.9a, 
the path is kept relatively straight during the relaxation. 
The modified NEB converges better than the regular NEB which has no perpendicular component of the 
spring force, as illustrated in Fig. B.8.
Fig. B.9b
shows that the reaction path obtained after 30 000 steps of the modified NEB rises to a significantly lower energy  
than the unconverged path given by the regular NEB method after the same number of iterations.

For such a long path, many images must be used in the NEB calculation in order to resolve all intermediate minima. The number of images can be increased by splitting the path in several parts. It is then possible to resolve elementary steps as is illustrated in a close-up of a short segment of the path
near the maximum, shown in Fig. 
B.10.

\begin{figure}
\includegraphics[width=0.8\linewidth]{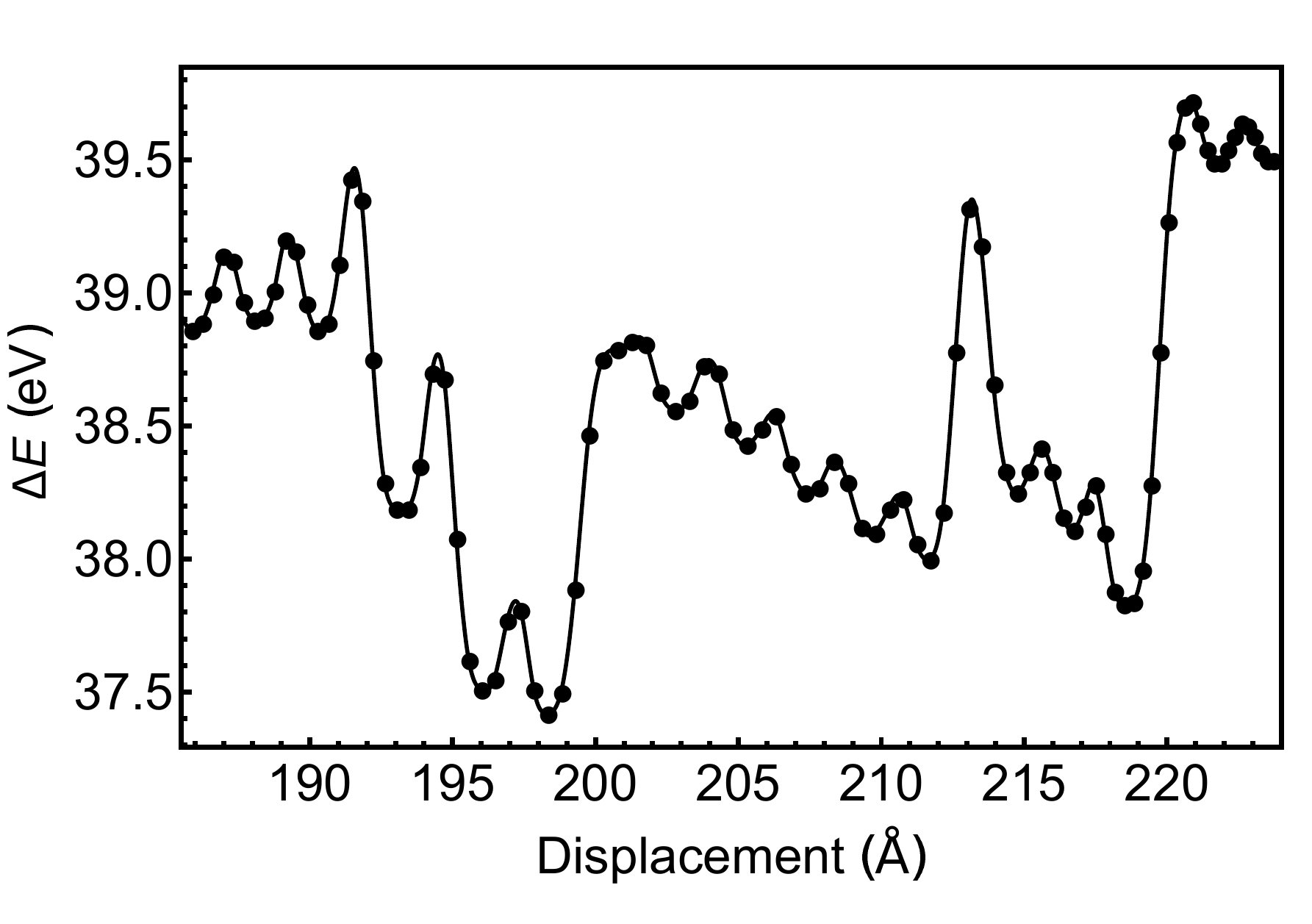}
\label{fig:NRJProfRefine}
\caption{
A subsection of the path shown in Fig. 
4 in the region around the maximum energy. A cubic spline interpolation is shown with a solid line.
The scale on the $x$-axis is the same in the two figures.  
Here, the energy barrier of each elementary step in the complex path can be seen. 
}
\end{figure}

In this study, all our NEB calculations are carried out with the modified NEB method until the maximum value of the norm of the $3N$-dimensional force vector of each image has dropped bellow 0.01 eV/\AA.


\bibliographystyle{ActaMatnew-2}


\end{document}